\documentstyle[11pt,newpasp,twoside,epsf]{article} 
\begin{document}
\title{Planet Detection Capabilities of the \emph{Eddington} Mission}
\author{Hans J. Deeg} 
\affil{Instituto de Astrof\'{\i}sica de Andaluc\'{\i}a, Granada, 
Spain; hdeeg@iaa.es\\Instituto de Astrof\'{\i}sica de Canarias, Tenerife, Spain}

\author{Keith Horne}
\affil{St.Andrews University, School of Physics and Astronomy, North 
Haugh, St. Andrews, Fife KY16 9SS, Scotland, UK}
\author{Fabio Favata and the \emph{Eddington} Science Team\footnotemark }
\affil{Astrophysics Division -- Space Science Department of ESA, 
ESTEC, Postbus 299, NL-2200 AG Noordwijk, The Netherlands}

\footnotetext{C. Aerts, E. Antonello, M. Badiali, C. Catala,
J. Christensen-Dalsgaard, A. Gimenez, M. Grenon, A. Penny, H. Rauer,
I.W. Roxburgh, J. Schneider, N.R. Waltham}

\begin{abstract}
\emph{Eddington} is a space mission for extrasolar planet finding and
for asteroseismic observations. It has been selected by ESA as an
F2/F3 reserve mission with a potential implementation in 2008-13. Here
we describe \emph{Eddington}'s capabilities to detect extrasolar planets,
with an emphasis on the detection of habitable
planets. Simulations covering the instrumental capabilities of
\emph{Eddington} and the stellar distributions in potential target fields
lead to predictions of about 10,000 planets of all sizes and
temperatures, and a few tens of terrestrial planets that are
potentially habitable. Implications of \emph{Eddington} for future larger
scale missions are briefly discussed.
\end{abstract}

\begin{figure}
\plottwo{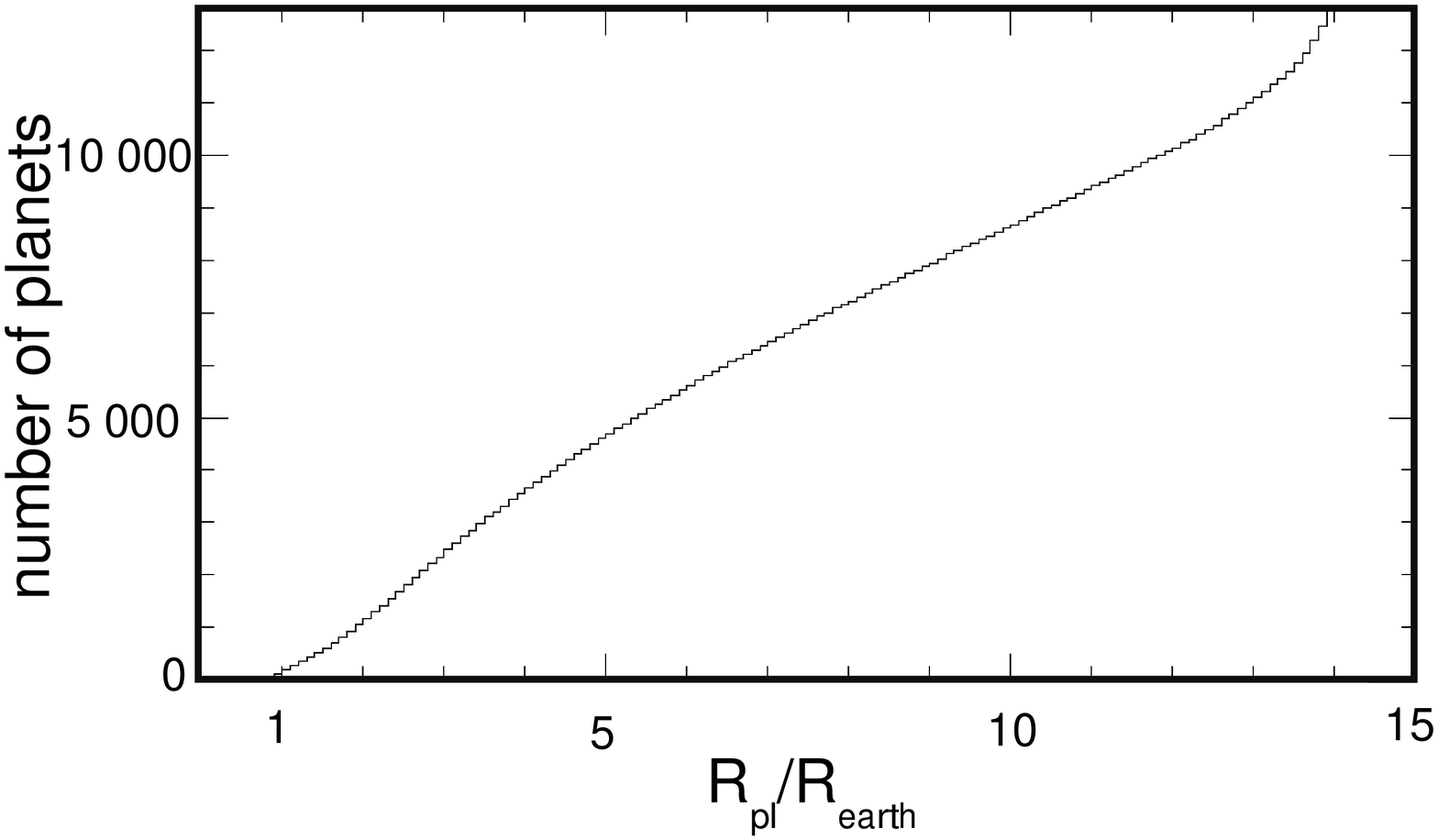}{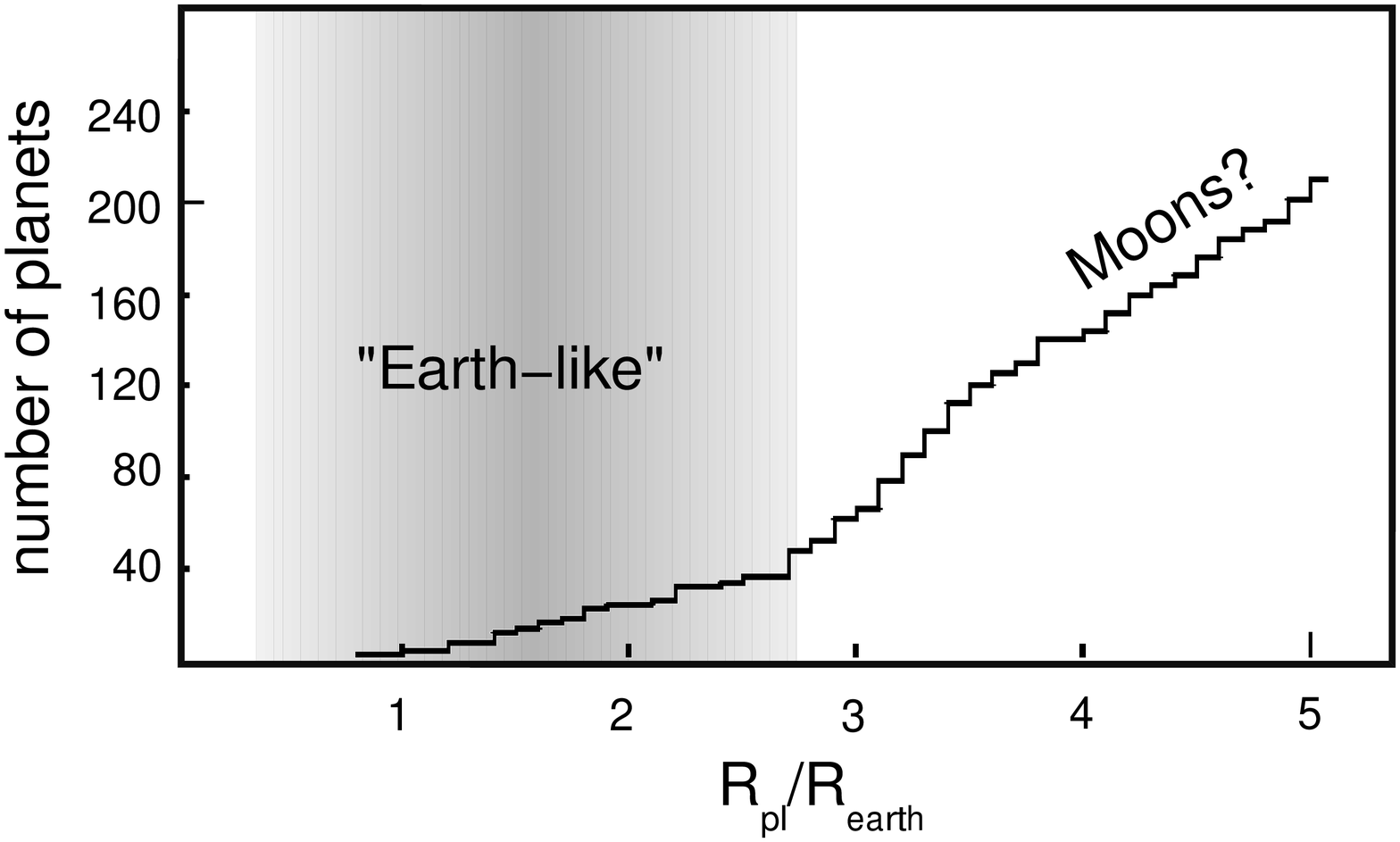} \caption{Expected count of
detected planets. Left: all planets at any temperature or distance
from the central star. Right: planets in the `habitable zone'
(250-350K). The field `Earth-like' indicates the size range of planets
suitable to support life (solid surface and ability to maintain
atmosphere). Moons around larger planets may also indicate life
sites.}
\end{figure}
\section{The \emph{Eddington} mission}
The \emph{Eddington} mission is a space telescope designed for two 
primary goals: asteroseismic studies and extrasolar planet finding. 
Both goals will be achieved through the acquisition of high-precision 
wide-field photometry, with a temporal stability that is possible only
from space. The basic design is an f/3 triple-reflecting telescope 
with a 1.2m diameter aperture, 0.6m$^2$ effective area,
and a 3$\deg$ diameter field of view imaged by a 20-CCD mosaic camera. 
The telescope will be launched into an orbit around 
the L2 Earth-Sun libration point, from which the entire sky is 
accessible for observations. The first 2 years of the mission will be 
dedicated primarily to asteroseismic studies, with pointings 
lasting up to two months to a variety of targets. The mission's second 
part will be dedicated to planet finding by the transit method, where 
the telescope will survey a single stellar field for 3 years. In 
total, about 500,000 stars will be surveyed by \emph{Eddington} as 
potential hosts for planetary systems.

In October 2000, \emph{Eddington} was selected by ESA as a reserve mission 
for the F2/F3 launch window (2009-2013). More details about the 
mission can be found in the \emph{Eddington} Assessment study (Favata et al., 
2000) and in the \emph{Eddington} web-site 
(http://astro.esa.int/SA-general/Projects/Eddington/). Here we will 
give a short overview about its planet detection capabilities.
\section{Planet detections}
The evaluation of \emph{Eddington}'s planet detection performance
consisted of two parts: First, we determined the kinds of planets this
mission can detect, in terms of the planet's size, orbital period, and
magnitude and type of central star (Deeg et al.\ 2000). Second, the
numbers and characteristics of the detected planets were estimated.
This employed Monte-Carlo simulations to fold \emph{Eddington}'s
`detection space' with models of the Galactic stellar distribution,
density and scale height (Robin \& Cr\'{e}z\'{e}, 1986). With
plausible exo-planet distributions, some 10,000 of planets of
\emph{any} size and orbital distance may be detected (Fig 1, left).
Of the detected planets the vast majority are both large and in close
orbits around their central stars. The most extreme ones, the `Hot
Giants', will be detected not only by transits, but also by the
reflected light of their central star (even if they do not produce
transits) yielding a large and nearly complete sample of these
planets. For smaller planets there are fewer sufficiently bright host
stars within the observational range, and they will be detected only
by transits. The habitable zone corresponds roughly to orbital
distances at which planetary surface temperatures of 250-350K are
expected. Under the assumption that every main-sequence star has one
planet in the habitable zone, a few hundred planets in that zone
should be detected, among them a few tens of habitable planets with
sizes of less then 2.5 $R_{Earth}$, which should have solid surfaces
(Fig 1, right). Around larger planets within the habitable zone there exists
the intriguing possibility that massive moons may be detected through
the perturbations they cause on the arrival time of transit minima
(Sarotetti \& Schneider, 1999). These moons may themselves
be harboring sites for life.

\begin{figure}
\plotone{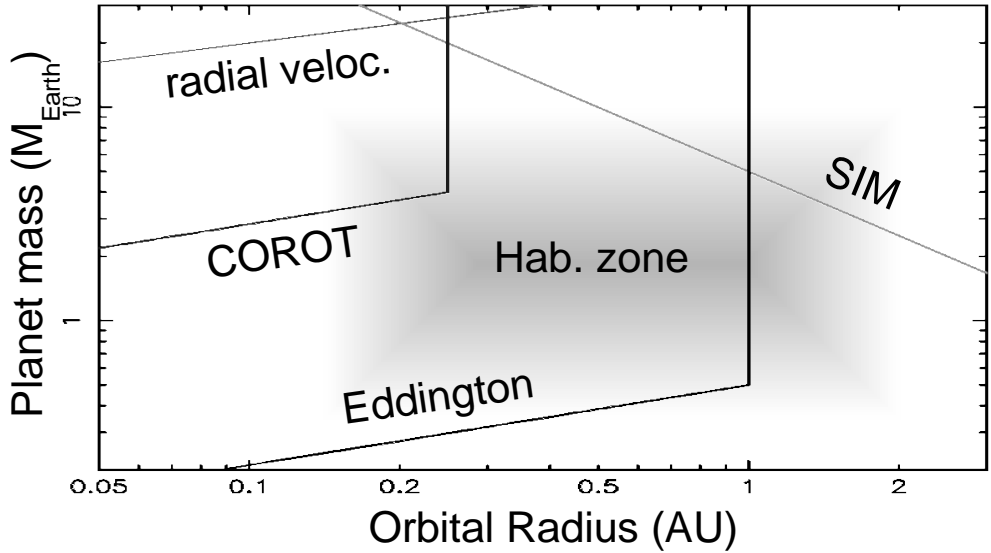} \caption{Comparison of planet finding 
capabilities of several missions. SIM's sensitivity is given 
for a star at 5pc distance. The limits for 
ground-based radial velocity searches are for a sensitivity of 1 m/s}
\end{figure}

\section{\emph{Eddington} in perspective}
As Fig.~2 shows, \emph{Eddington} is the first mission that will
survey the habitable zone, and deliver a broad overview
of the abundance of habitable planets in the Galaxy.
This is an important step beyond
missions like COROT and SIM, which may detect a few special cases of
habitable planets.
\emph{Eddington} will deliver a large well-defined statistical sample
of planets with known planet radii, orbital distances, 
and host star surface temperature and type.
This is critical input for the design and target selection
of future missions like Darwin and TPF that aim for detailed
characterizations of Earth-like planets.
\emph{Eddington} moves us one step closer to resolving the
age-old question: Are there other worlds like ours?


\begin{references}

\reference Deeg H.J., Favata F., and Eddington Science Team 2000, in
'Disks, Planetesimals and Planets', Tenerife,
Eds. F. Garzon et al., ASP Conf proc., in print. Preprint astro-ph/0006130



\reference Favata F., Roxburgh I., Christensen-Dalsgaard J. (Eds.) 
2000, 'Eddington Assessment Study Report', ESA report ESA-SCI(2000)8



\reference Robin A., Cr\'{e}z\'{e} M. 1986, A\&A 157, 71
\reference Sarotetti P., Schneider J. 1999, A\&AS 234, 553

\end{references}
\end{document}